\documentclass{jfm}

\usepackage{graphicx}
\usepackage{epstopdf,epsfig}
\AppendGraphicsExtensions{.tif}
\AppendGraphicsExtensions{.tiff}

\usepackage{newtxtext}
\usepackage{newtxmath}
\usepackage{natbib}
\usepackage{hyperref}
\hypersetup{
    colorlinks = true,
    urlcolor   = blue,
    citecolor  = black,
}

\newcommand{\RomanNumeralCaps}[1]
\linenumbers

\newcommand\gi{\mathrm{g}}

\title{Attenuating surface gravity waves by an array of submerged resonators: an experimental study}

\author{
  Matteo Lorenzo              \aff{1},
  Paolo Pezzutto              \aff{2} \corresp{\email{paolo.pezzutto@cnr.it}},
  Filippo De Lillo            \aff{1},
  Francesco Michele Ventrella \aff{1},
  Francesco De Vita           \aff{3},
  Federico Bosia              \aff{4}
  \and
  Miguel Onorato              \aff{1}
  }

\affiliation{\aff{1}Dipartimento di Fisica and INFN, Università di Torino, Via P. Giuria 1, 10125 Torino, Italy
\aff{2}Istituto per le Risorse Biologiche e le Biotecnologie Marine, CNR, Largo Fiera della Pesca 2, 60125 Ancona, Italy
\aff{3}Dipartimento di Meccanica, Matematica e Management (DMMM), Politecnico di Bari, Via Orabona 4, 70125 Bari, Italy
\aff{4}Dipartimento Scienza Applicata e Tecnlogia (DISAT), Politecnico di Torino, Corso Duca degli Abruzzi 24, 10129 Torino, Italy}

\begin{document}
\maketitle
\begin{abstract}

We report on an experimental study of a device composed by an array of submerged, reversed and periodic cylindrical pendula (resonators), whose objective is the attenuation of surface gravity waves. The idea is inspired by the concept of metamaterials, \emph{i.e.} engineered structures designed to interact with waves and manipulate their propagation properties. The study is performed in a wave flume where single frequency waves are excited in a wide range of frequencies. We explore various configurations of the device,  measuring the transmitted, reflected and dissipated energy of the waves. If the incoming wave frequencies are sufficiently close to the natural frequency of the pendula, we find a considerable wave attenuation effect, driven by viscous dissipative mechanisms. This behaviour is enhanced by the number of resonators in the array. Moreover, the device is also capable of reflecting the energy of selected frequencies of the incoming waves. These frequencies can be predicted by assuming the interactions involving at least three wave modes, including higher harmonics, and are therefore associated with the distance between the resonators. The presented results show promise for the development of a environmentally sustainable device for mitigating waves in coastal zones.

\end{abstract}




\section{Introduction}
\label{sec:intro}
It is well established that anthropogenic pressure, combined with natural processes, has contributed both to a worsening of the environmental quality of coastal areas and to the triggering of erosion dynamics, resulting in the instability of rocky coasts and the retreat of sandy ones \citep{Athanasiou2020}.
The most important natural factors responsible for coastal erosion are wind, wave motion, currents, lack of sediments from rivers into the seas and movements of the soil \citep{OppenheimerIPCC2019}. The main anthropogenic factors are mainly linked to the construction of infrastructures and residential and industrial settlements.
According to \citet{Luijendijk2018} over 44\% of the world’s sandy beaches are persistently eroding, with many of these beaches situated in Europe. On the global scale, the average erosion rate for the period 1984-2016 has been at least 1 m per year \citep{Luijendijk2018}.

Traditionally, the most effective solutions used to counteract beach erosion include groins and breakwaters \citep{Hughes1993}. Groins are shore perpendicular structures that are meant to capture sand transported by the long-shore currents. One of the main disadvantages of groins is that debris may accumulate around them, creating problems for marine animals. Besides, they constitute a foreign element in the coastal landscape due to their unnatural shape perpendicular to the shoreline. Breakwaters are structures built parallel to the coast that reduce the intensity of wave action in inshore waters. They are placed 30 to 90 m offshore in relatively shallow water. Breakwaters are meant to determine the line where the waves break and tend to move the coastal currents towards the open sea. This creates a discontinuity in the solid transport with a consequent reduction in the contribution of sediments from the protected beach to the neighboring coasts. Frequently, such structures lead to a deterioration in the quality of the waters with an increase in the turbidity and impact on the colonisation of marine organisms \citep{FIRTH2014122}. Furthermore, wave motion at the head of these structures magnifies scours, deepening the seabed, and induces rip currents, creating dangers for bathing.
One should also consider the fact that such hard-engineered structures will need to cope with sea level rise \citep[see e.g.][]{Pranzini2015}. In many situations, to preserve the equilibrium of already protected beaches, there will be the need to adapt or totally rebuild the existing structures. There is therefore the need to explore the feasibility of alternative solutions and complementary strategies, which can be more adaptable and less impacting than the state of the art.
Tethered  breakwaters, built as regular lattices of floating structures, can provide an attractive option for beach management \citep{Dai2018}. Compared to rubble mound breakwaters, this type of structure has a limited impact on water circulation. Without localized refraction/diffraction effects, the formation of tombolos and stagnating pools between them is prevented. Moreover, if the floaters are kept below the water surface, their visual impact from the shore is negligible, similar to artificial reefs \citep{McCartney1985, Dai2018}. In the presence of rising sea levels, tethered floats are also more adaptable. The environmental impact of point-wise hinged floaters is very small throughout the life of the structure, from deployment to decommissioning. Depending on their efficiency, the cost-effectiveness of beach protection using tethered floats is favourable compared to the current hard-engineering strategies.
Tethered floating breakwaters have been studied since the late 70s with extensive laboratory and field tests, exploring different configurations for the anchoring and submergence of the floats \citep[see][]{Essoglou1975, Agerton1976, Seymour1976, Jones1978}. The models proposed for the efficiency of these devices are based on experimental results, obtained with waves impacting on a single float (\textit{i.e.} single row of floats) \citep{Agerton1976, Seymour1979}. The underlying hypothesis is that subsequent floats behave with the same efficiency. For this reason, fluid drag dissipation is the only wave attenuation mechanism that has been considered so far, while wave scattering has been considered negligible (\cite{Seymour1979}). Wave reflection from a single float (or a single row of floats) might be small \citep{Dean1948, Chaplin1984, EvansLinton1989, Grue1992}; however, in principle, we cannot discard the hypothesis that multiple float systems might respond to waves with some collective behavior, favouring the wave scattering.

In this paper,  we will consider an array of tethered submerged  floaters, with a periodic configuration, inspired by the concept of metamaterial wave control. Metamaterials are engineered structures, often distributed in periodic patterns, designed to interact with waves and manipulate their propagation properties, such as phase and group velocity, or to produce effects such as negative refraction, cloaking, superlensing and absorption \citep{Hussein2014}. In terms of wave attenuation, the key concept is that the metamaterial dispersion relation may be non-monotonic or even involve \emph{band-gaps}, i.e. ranges of frequencies where wave propagation is inhibited \citep{Laude2015}. The metamaterial concept was first developed in
the field of optics \citep{Pendry2001} and later extended to phononic crystals and elastic waves \citep{Hussein2014, Laude2015}. Nowadays, they find applications in different fields of physics and engineering, from seismic protection \citep{Brule2020} to non-destructive testing \citep{Miniaci2007}.

So far, the number of results involving metamaterials in the field of water waves is very limited due to the inherent complexity of the problem, which involves the coupling of oscillating bodies inside a fluid, with correlated flow, surface and viscous effects \citep{DeVita2021sim}. Numerical models of periodic vertical cylinders \citep{Hu2005, Zheng2020} have been implemented, showing results in terms of refraction, amplification or rainbow reflection \citep{bennetts2018graded, archer2020experimental, wilks2022rainbow}.  \citet{Hu2011g} showed that propagation of water waves through a periodic array of resonators is inhibited near a low resonant frequency, as if water had a negative effective gravity.

Studying a periodically drilled bottom, \citet{Hu2003} discussed both band-gap formation and manipulation. Other works considered the interaction of gravity waves with a macroscopic periodic structure, like a sinusoidal floor \citep[see][]{Davies1984, Hara1987, Kar2020}, showing the presence of Bragg scattering mechanisms and the existence of a band-gap structure determined by Bragg's law. Bragg scattering was also shown to be induced by a train of fixed floating pontoon breakwaters \citep{Ouyang2015}.

It is worth mentioning that our device is similar to wave energy conversion systems based on single point-absorbers, on which extensive literature can be found \citep[see \emph{e.g.}][]{Evans1979, Crowley2013, Anbarsooz2014, Sergiienko2017, Dafnakis2020}. Our purpose here is to investigate to which extent the concept of metamaterial wave control can be applied to an efficient attenuation system for surface gravity waves, rather than focusing on wave energy extraction. The periodic structure is designed as a lattice of submerged inverted pendula, in which each pendulum is anchored to the sea bed. Using direct numerical simulations of the incompressible Navier-Stokes equation in its two-dimensional form with periodic boundary conditions and moving bodies, \citet{DeVita2021meta} showed that large energy attenuation is possible if the wave and the pendula resonances are characterized by similar frequencies (see \citet{DeVita2021sim} for computational details). As the number of resonators is increased, the range of attenuated frequencies also increases. Here, we consider laboratory experiments aiming to characterize the dissipation and reflection processes, looking for a possible collective behaviour.

The paper is organized as follows. Section \ref{sec:lab_tests} describes the metamaterial device, the experimental setup and the analysis techniques. Section \ref{sec:res} discusses the experimental results, summarizing them in terms of the capability of the metamaterial to dissipate and scatter the incoming wave energy. The net effects in terms of wave energy reduction are also reported. Section \ref{sec:discussion} contains the interpretation of the device behavior, linking the single mechanisms to the metamaterial structure. Finally, in section \ref{sec:concl} conclusions and some future possible developments are discussed.

\section{Methodology}

\label{sec:lab_tests}

\subsection{The experimental setup}

\begin{figure}
  \centerline{\includegraphics[width=0.95\textwidth]{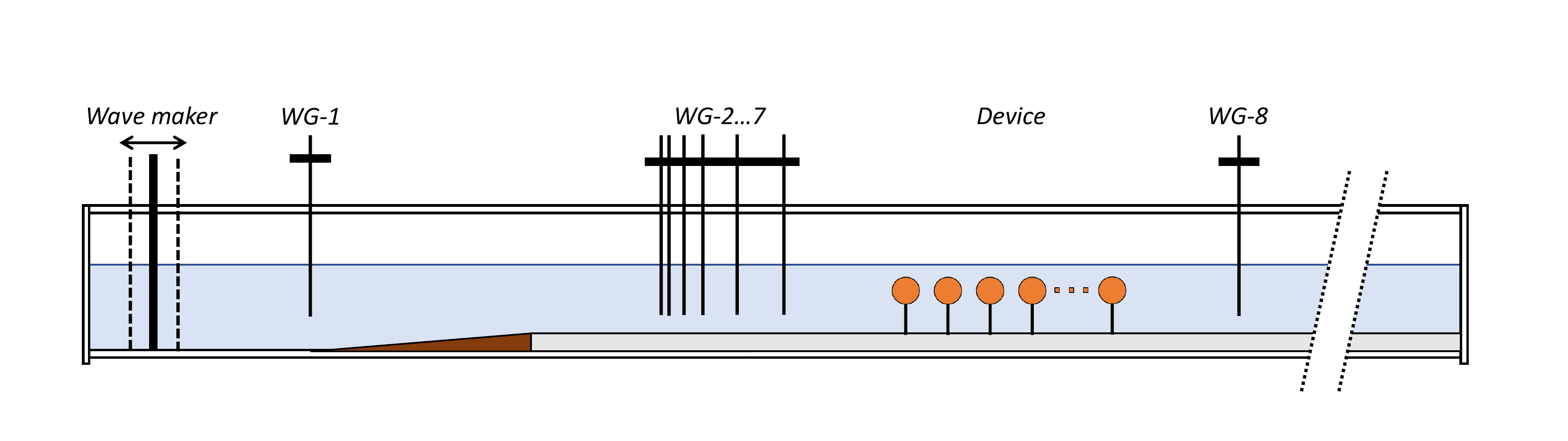}}
  \caption{Sketch of the testing facility, experimental set-up and physical model (not in scale). The distance  of the different wave gauges with respect to the wavemaker at rest position are listed in table \ref{tab:expsetup_distances}.}
\label{fig:waveflume}
\end{figure}

Experiments were performed at the “Giorgio Bidone” Hydraulics and Fluid Dynamics Laboratory of the Politecnico di Torino, in a 50 m long, 0.6 m wide and 1 m deep wave flume (Figure \ref{fig:waveflume}).
The sidewalls are made of glass for the entire wave flume length. The flume bottom is made of concrete, whose Nikuradse equivalent sand roughness $k_s$ is about 0.25 mm \citep{Peruzzi2020}. In the wave generation zone, the flume is 0.1 m deeper, and a stainless-steel ramp  (1.6 m long, located at 2.1 m from the wavemaker rest position), which has been designed to prevent boundary layer separation \citep{Peruzzi2020}, links this zone with the main channel. A graded rod enables still water depth $h$ measurement with 0.05 mm accuracy.

Waves were generated by a piston-type wavemaker (made by Delft Hydraulics Laboratory), driven by an electric motor. The input signal for regular waves was generated by in-house MATLAB-based scripts. The paddle is provided with 3 metallic plates, in order to prevent the excitation of spurious transverse modes.

\begin{figure}
  \centerline{\includegraphics[width=0.95\textwidth]{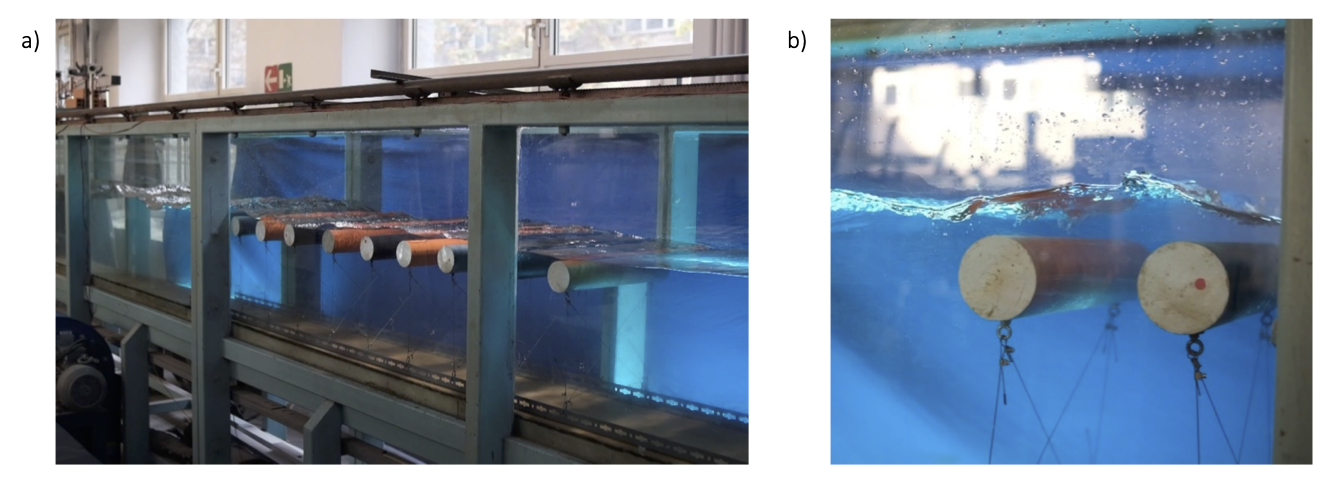}}
  \caption{a) Snapshot of 11 cylinders device forced by regular waves of 2 cm amplitude and 1.08 Hz frequency, moving left to right. On the left side of the device, superimpositions of incident and reflected waves are visible. On the right, the wave attenuation can be observed. b) Zoom on the first two cylinders. }
\label{fig:device_scheme}
\end{figure}

The wave attenuation device consisted of an array of submerged floating cylinders (see Figure \ref{fig:device_scheme}), placed with their 58 cm axes parallel to the wave crests. The distance between the bottom of the flume and the top of the cylinder was set equal to 43 cm. Cylinders were made with commercial PVC pipes (with a diameter of 82 mm, and wall thickness of 3 mm) filled with air.
The tube ends were closed with 20 mm thick discs made of polyurethane foam and sealed with silicone caulk. Each cylinder was moored to two matching perforated steel angle bars (20 mm $\times$ 20 mm $\times$ 2 mm), which were fixed to the concrete blocks on the flume bottom, (see figure \ref{fig:device_scheme}), starting at 15.7 m from the wavemaker rest position. Each bar was 4 m long, with a series of holes 11.75 cm apart (indicated in the following as the $L^*$ parameter). The anchoring of each cylinder was made with four co-planar cables (made of galvanized steel with a diameter of 1.2 mm) connecting each tube end with both angle bars. This allowed to prevent transverse horizontal motion.

The water surface elevation was measured with a commercial system (WG8USB, Edinburgh Designs Ltd.) that is able to sample the surface displacement with an accuracy of $\pm$0.05 mm at a rate of 128 Hz via eight 700 mm-long, resistance-type wave gauges. One wave gauge (WG1) was placed in the flume wave generation zone to monitor the wavemaker-generated waves. Another single gauge (WG8) was placed on the lee side of the device mounting zone to obtain measurements of the transmitted wave field. The remaining six gauges (WG2-WG7) were grouped into an array placed on the sea side of the device, with the last gauge located at 2.32 m from the first cylinder anchoring position. All the gauge positions  (which are listed in table \ref{tab:expsetup_distances}) were carefully chosen to be sufficiently far from any obstacle, to prevent the capturing of localized evanescent modes. We used the gauge array (WG2-WG7) to obtain, via a least squares scheme \citep{Zelt1993}, the linear incident and reflected surface elevation on the sea side of the device.

\begin{table}
  \begin{center}
\def~{\hphantom{0}}
  \begin{tabular}{ccccccccc}
      $\;$ & WG1$\;$ & WG2$\;$ & WG3$\;$ & WG4$\;$ & WG5$\;$ & WG6$\;$ & WG7$\;$ & WG8$\;$ \\
      Distances [m]$\;$ & 1.80$\;$ & 12.26$\;$ & 12.36$\;$ & 12.51$\;$ & 12.72$\;$ &  13.02$\;$ &  13.47$\;$ &  21.10$\;$
  \end{tabular}
  \caption{Distances between the wavemaker rest position and the i-th wave gauge (WGi).}
  \label{tab:expsetup_distances}
  \end{center}
\end{table}

\subsection{Experimental conditions}

The device was composed of a number of cylindrical pendula placed at a regular distance, so as to maintain the periodic structure (see figure \ref{fig:device_scheme}). The number of cylinders was tentatively designed to cover one wave length corresponding to the natural frequency of the single pendulum (see \ref{sub:freq_pend}), while maintaining a mutual distance $L$ sufficient enough to avoid possible direct interactions between two consecutive pendula. The 11-cylinder configuration can thus be thought as the smallest representative realization of a metamaterial, whose properties are usually studied for infinitely long arrays.

The test program (table \ref{tab:exp_test}) included various geometrical configurations in terms of number of cylinders $N$, spacing between cylinders $L$, water depth $h$ and wave amplitude $a$. All mutual distances $L$ were multiples of the distance $L^*$ between adjacent holes in the perforated bars.
With the aim of characterizing the frequency response of the device, each configuration was tested with about 30 regular long-crested wave conditions, with carrier frequency $f$ spanning between 0.4 and 1.4 Hz. Each single run was programmed to be 80 s long, including 3 s of initial ramp, starting at still water condition, and the gauge acquisition duration was set to 160 s, starting 5 s before the onset of wave generation.
A fundamental constraint for the design of this wave attenuator is that the floaters must remain below the surface, at least in mild sea conditions. The water depth considered in most of the runs was  $h=$ 0.45 m, while the top of the pendula was set to 0.43 m for all experiments. In this conditions, the submergence of the pendula (measured from the still water level and the top of the cylinder) measures $d=$ 2 cm. Forcing the system with waves whose amplitude is 1 cm, the 2 cm gap is enough to keep the pendula submerged, even during the passage of the incident wave.

\begin{table}
  \begin{center}
\def~{\hphantom{0}}
  \begin{tabular}{cccc|cccc}
      $\;N$ & $h\;$[m]  & $a\;$[m] &  $L\;$ &  $\;N$ & $h\;$[m] & $a\;$[m] &  $L\;$ \\
      $\;$2 &     0.45  &    0.01  & 2$L^*$ &  $\;$5 &     0.45 &     0.02 & 4$L^*$ \\
      $\;$2 &     0.45  &    0.02  & 2$L^*$ &  $\;$5 &     0.45 &     0.01 & 7$L^*$ \\
      $\;$2 &     0.45  &    0.01  & 4$L^*$ &  $\;$5 &     0.45 &     0.02 & 7$L^*$ \\
      $\;$2 &     0.45  &    0.02  & 4$L^*$ & $\;$11 &     0.43 &     0.01 & 2$L^*$ \\
      $\;$2 &     0.45  &    0.01  & 7$L^*$ & $\;$11 &     0.45 &     0.01 & 2$L^*$ \\
      $\;$2 &     0.45  &    0.02  & 7$L^*$ & $\;$11 &     0.47 &     0.01 & 2$L^*$ \\
      $\;$3 &     0.45  &    0.01  & 2$L^*$ & $\;$11 &     0.49 &     0.01 & 2$L^*$ \\
      $\;$3 &     0.45  &    0.02  & 2$L^*$ & $\;$11 &     0.45 &     0.02 & 2$L^*$ \\
      $\;$3 &     0.45  &    0.01  & 4$L^*$ & $\;$11 &     0.45 &     0.03 & 2$L^*$ \\
      $\;$3 &     0.45  &    0.02  & 4$L^*$ & $\;$11 &     0.43 &     0.01 & 3$L^*$ \\
      $\;$3 &     0.45  &    0.01  & 7$L^*$ & $\;$11 &     0.45 &     0.01 & 3$L^*$ \\
      $\;$3 &     0.45  &    0.02  & 7$L^*$ & $\;$11 &     0.47 &     0.01 & 3$L^*$ \\
      $\;$4 &     0.45  &    0.01  & 2$L^*$ & $\;$11 &     0.49 &     0.01 & 3$L^*$ \\
      $\;$4 &     0.45  &    0.02  & 2$L^*$ & $\;$11 &     0.45 &     0.02 & 3$L^*$ \\
      $\;$4 &     0.45  &    0.01  & 4$L^*$ &                                       \\
      $\;$4 &     0.45  &    0.02  & 4$L^*$ &                                       \\
      $\;$4 &     0.45  &    0.01  & 7$L^*$ &                                       \\
      $\;$4 &     0.45  &    0.02  & 7$L^*$ &                                       \\
  \end{tabular}
  \caption{Summary of the tests conducted in the lab. Each row represents a configuration which is defined by the number of pendula, $N$, the water depth ($h$), the wave amplitude ($a$) and the spacing between two consecutive pendula, $L$. The latter is a multiple of  $L^*=11.75$ cm. Each configuration has been tested under the same regular sea conditions, for frequencies spanning the range $0.4$ to $1.4$ Hz.
  }
  \label{tab:exp_test}
  \end{center}
\end{table}

\subsection{Wave analysis}
\label{sub:wave_analysis}
\begin{figure}
  \centerline{\includegraphics[]{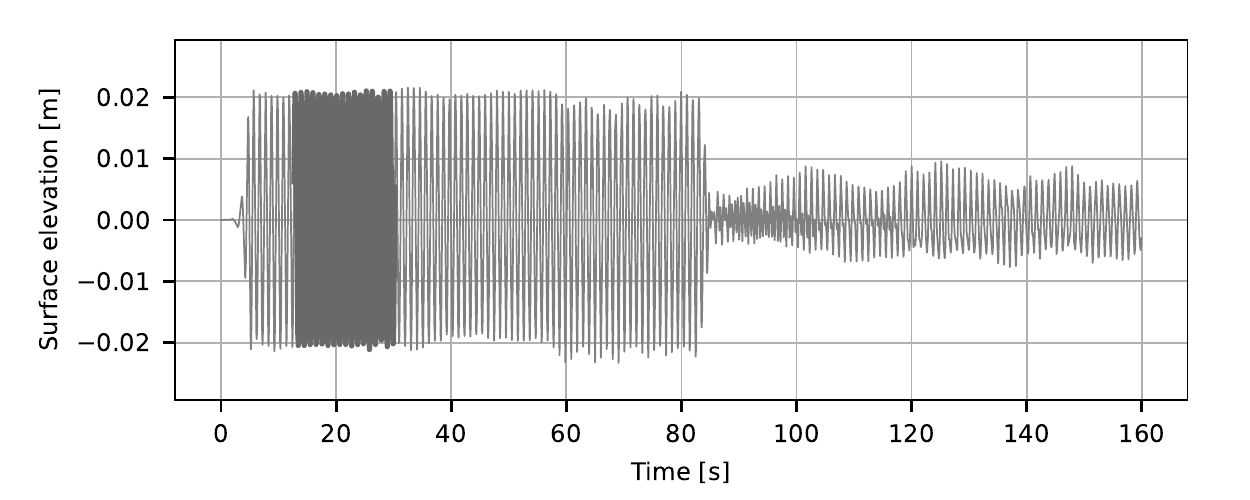}}
  \caption{Example of surface elevation time series acquired on the sea side of the pendula array (solid light grey line). The solid dark line highlights the stationary time-series used to assess the main properties.}
\label{fig:ex_allTS}
\end{figure}

There is a number of features that are common to all test signals, as can be seen in the example shown in Fig. \ref{fig:ex_allTS}. After the initial rest phase, there is a wavemaker ramp-up of a few oscillations, followed by a further two or three wave periods for the system to reach a stationary state. After a stationary phase, the state is perturbed by reflected waves coming from either the wavemaker or the flume end.

In all experiments, only the stationary signal was considered for data processing, excluding all transient states. Thus, in each test, a different number of waves was analysed, depending on the associated wavelength, the truncated signal length being the same for incident, reflected and transmitted time-series. In any case, the maximum possible number of wave periods was considered, allowing a more robust estimation of the desired quantities at higher frequencies.

While the transmitted signal was simply measured using a single gauge (WG8) placed on the shore side of the device, incident and reflected surface elevation were extracted by processing all signals measured by the 6 gauge array, \emph{i.e.} WG2-WG7 located on the offshore side of the device.
The incident-reflected separation approach is based on the best-practice least squares scheme proposed by \citet{Zelt1993}, which considers that only free progressive modes exist in the flume. 
As can be deduced from Table \ref{tab:exp_test}, according to weakly nonlinear theory \citep[see \emph{e.g.}][]{Whi74} the amplitude of the second order bound modes in the channel were at most $1\%$ of the amplitude of the carrier wave, so the overall error in estimating them as free modes is negligible. Moreover, although evanescent modes must be expected close to the obstacles, their measurement was avoided by placing all gauges sufficiently far from the obstacles.
A purely linear approach was thus sufficient to analyse the stationary parts of the signals, but, given the presence of transients (figure \ref{fig:ex_allTS}), the application of the least squares scheme to the whole time series would introduce nonphysical artefacts. To avoid this, localized and overlapped portions of the signals were processed using a sliding Hanning window. The window length was designed to approximately equal an integer number of periods, also oversampling the signal according to a first estimation of the main wave period. Moreover, the start and end of the series were artificially extended with additional zeros, to sufficiently resolve the frequency domain, and thus correctly estimate the wavenumbers.

Once the stationary portions of the incident, reflected and transmitted surface elevation were obtained, we estimated the reflection and the transmission coefficients $C_R$ and $C_T$ as ratios between energies (spectral zero-th moments $m_0$). From energy balance, we derive the dissipation coefficient as

\begin{equation}\label{eq:equilibrium}
  C_D = 1 - C_R - C_T.
\end{equation}

\begin{figure}
  \centerline{\includegraphics[]{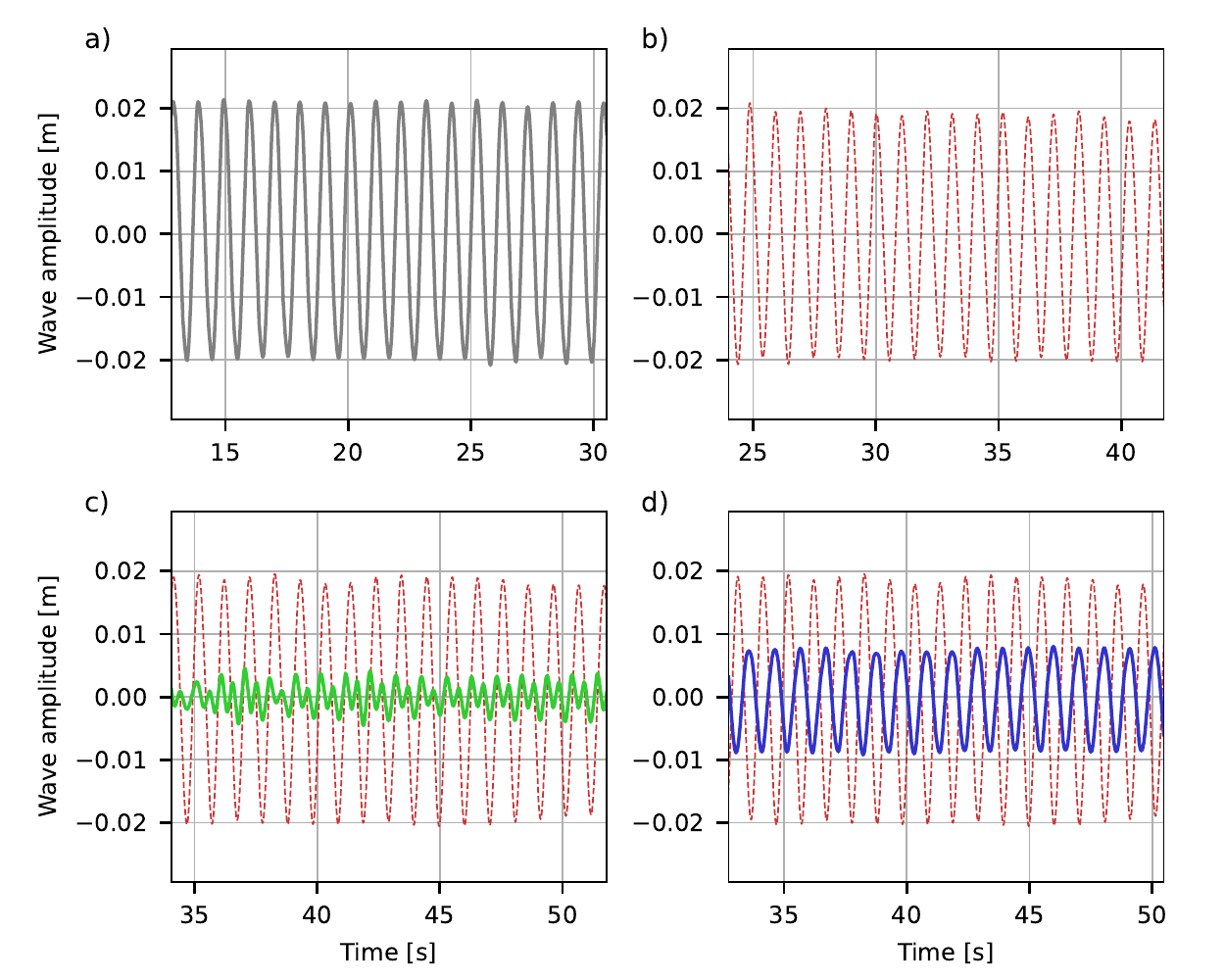}}
  \caption{
  Example of generated (a), incident (b), reflected (c) and transmitted (d) waves. Subplots have different abscissae corresponding to different transit times of the same wave crest. In panels (c) and (d), the background incident wave field (dotted-red line) has been included for visual comparison.}
\label{fig:ex_cuttedTS}
\end{figure}

Figure \ref{fig:ex_cuttedTS} shows an example of a time series for a wave with an amplitude of 2 cm and frequency of 0.97 Hz interacting with 11 cylinders moored with $2L^*$ mutual spacing. The grey-scale time-series (figure \ref{fig:ex_cuttedTS}a) represents the real signal acquired by WG1. The red line, instead (figure \ref{fig:ex_cuttedTS}b, also repeated in figures \ref{fig:ex_cuttedTS}c and \ref{fig:ex_cuttedTS}d), shows the incoming series, including the waves reflected by the device (green signal in figure \ref{fig:ex_cuttedTS}c) and the transmitted waves at the shore-side of the device (blue time series in \ref{fig:ex_cuttedTS}d).
\subsection{Pendulum frequency estimation}
\label{sub:freq_pend}
Pendula were tested in calm water to estimate their natural (resonance) frequency, $f_r$. A cylinder was first slightly dislocated from its rest position, kept in position for a couple of seconds, and then released as instantaneously as possible. By means of a stopwatch, we measured the time interval between the second and the fifth oscillations. To reduce the uncertainty of the result, this simple test was repeated for at least ten times per cylinder, and performed on all cylinders, which differed only by a few grams. A common value of 0.6 Hz was found. All the results used in this section are obtained by averaging several cylinders.

Starting from the pendula resonant frequency, the added mass coefficient can be estimated \citep{Neill2007}. Indeed, when a body of mass $M$ in a fluid is subjected to an acceleration $a$, the fluid around the body is also accelerated. Thus, the total kinetic energy is given by two terms and the total force $F$ will be
\begin{equation}
    \label{eq:total_force}
    F = \left(M+m\right) a,
\end{equation}
where $m$ is the so-called \textit{added mass}. The latter depends on the volume $V$ of fluid displaced, and consequently on the shape of the body itself, as $c \rho_w V$, with $c$ a coefficient depending on the shape of the body.
\begin{figure}
  \centerline{\includegraphics[width=0.8\textwidth]{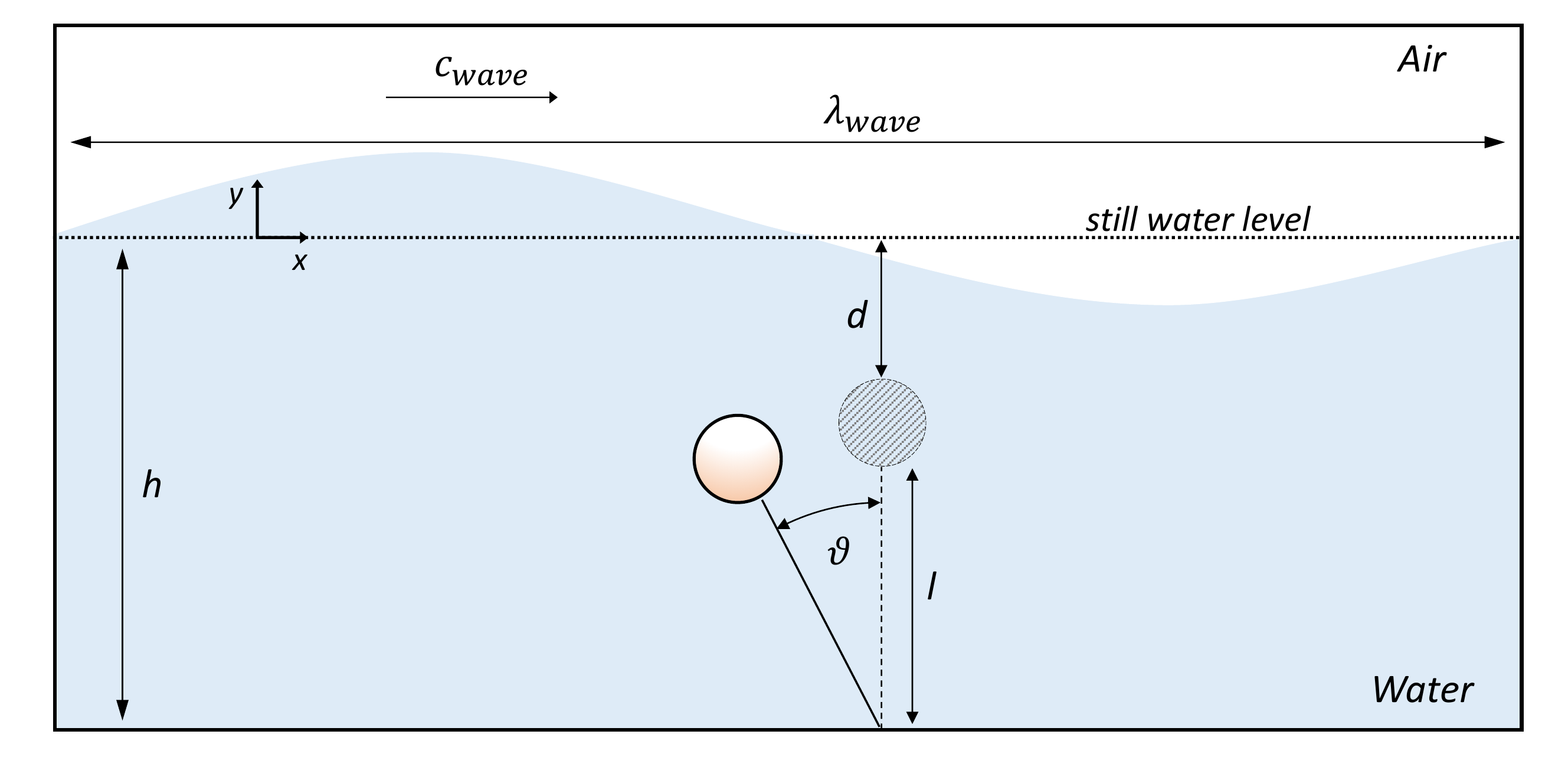}}
  \caption{Schematic of a single pendulum oscillating under the action of a regular wave, moving forward in the x direction. The vertical pendulum position shows the rest condition.}
\label{fig:pendulum_scheme}
\end{figure}
The reversed pendulum under the water surface (sketched in figure \ref{fig:pendulum_scheme}) was composed by a wire of length $l$ = 0.37 m and a cylinder of density $\rho_p$ = 220.27 kg/m$^3$ (estimated from a total measured mass of 0.7 kg and a volume $V_p$ of 3168 cm$^3$), smaller than that of the surrounding water $\rho_w$ = 997 kg/m$^3$. The forces acting on the pendulum are weight and buoyancy. Indicating with $\theta$ the angle between the pendulum wire and the vertical axis, the equation of motion can be written as
\begin{equation}
    \frac{d^2\theta}{d t^2} = - \frac{l \; sin\theta \; g  (\rho_w V - \rho_p V)}{\mathcal{I}} ,
\end{equation}
where $\mathcal{I} = M l^2$ is the moment of inertia, $M=\rho_p V$ is the mass of the cylinder and $g$ the gravitational acceleration. Considering small angle oscillations, an equation for an harmonic oscillator can be obtained and the consequent expression for the angular frequency is

\begin{equation}
    \label{fig:omega_p1}
    \omega = \omega_a \; \sqrt{\frac{\rho_w - \rho_p}{ \rho_p}},
\end{equation}

where $\omega_a = \sqrt{g/l}$ is the angular frequency of an oscillator in vacuum. As mentioned, the total inertial mass within the fluid can be defined as

\begin{equation}
    \label{fig:totalmass}
    m_i = M + c \rho_w  V,
\end{equation}

Considering the added mass, the pendulum frequency \ref{fig:omega_p1} is modified to
\begin{equation}
    \label{fig:omega_p2}
    \omega = \omega_a \; \sqrt{\frac{\rho_w - \rho_p}{ \rho_p + c \rho_w}}.
\end{equation}
Thus, while the oscillation period of a simple pendulum in vacuum is independent of its mass, this is no longer true in a fluid, were both buoyancy and added mass play a role.
Solving the equation \ref{fig:omega_p2} for $c$ and substituting all the known
values, we obtained an experimental added mass coefficient $c \approx 1.25$.

\section{Results}
\label{sec:res}
\subsection{Dissipation}
\label{sub:dissip}

\begin{figure}
 \centerline{\includegraphics[]{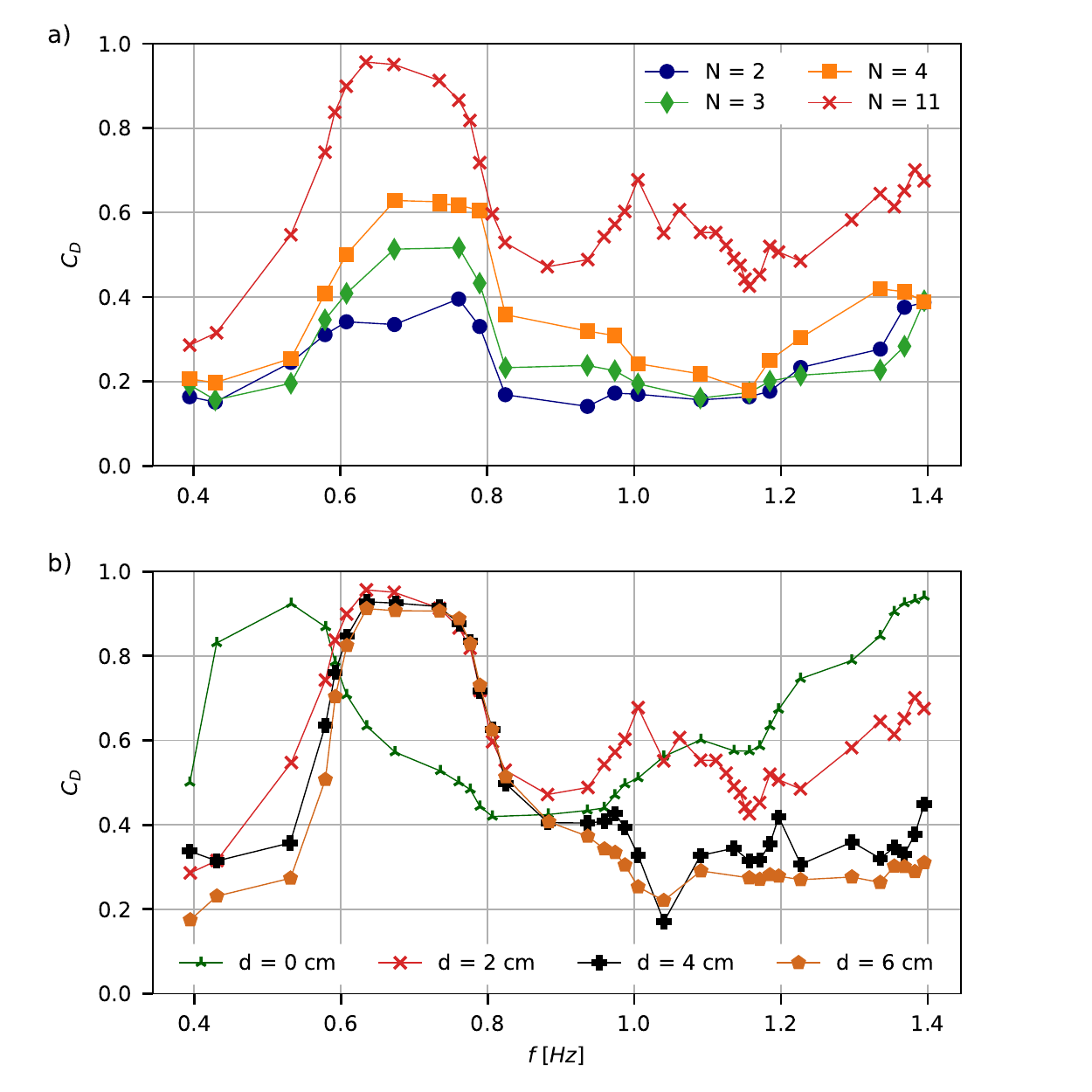}}
  \caption{Dissipation coefficient as function of the generated wave frequency.
  a) Each series represents a configuration with a different number of resonators, for L = 0.24 m, a = 0.01 m, h = 0.45 m.
  b) Each series represents a different water level, for $L = 2L^*$, a = 0.01 m, N = 11 cylinders.
The red-cross line is the same in both plots}
\label{fig:dissip_11cil}
\end{figure}

In figure \ref{fig:dissip_11cil}, the dissipation coefficient, computed according to equation \eqref{eq:equilibrium}, is plotted against the frequency of the incident waves. In the first panel, figure \ref{fig:dissip_11cil}a), each curve represents a configuration with the same mutual distance between the pendula. \textit{i.e.} $L=2L^*$, the same water depth, but a varying number of pendula.
Regardless the number of cylinders, high energy losses occurred in the band $0.6-0.8 $ Hz, with slightly larger maxima than $f_r=0.6$ Hz. When increasing the number of pendula (up to 11), the dissipation increases, especially in the band $0.6-0.8$ Hz.

In figure \ref{fig:dissip_11cil}b, we considered the effect of the distance $d$ between the still water level and the top of the cylinders on the dissipation. Each curve represents a different submergence while the device was fixed on a 11 cylinder configuration. Given the wave amplitude $a=1$ cm, for three out of four water levels, cylinders remained submerged both at rest and during the passage of the wave troughs. In these three cases, \emph{i.e.} $h\geq45$ cm (or equivalently $d\geq2$ cm), large dissipation occurs in the same $0.6-0.8$ Hz band. The behavior is different in the case of lower water level ($h=43 $ cm) corresponding to the slight emergence of the cylinders during the passage of the wave troughs. The maximum of energy dissipation occurs at lower frequencies, below the modal one.
In the high frequency tail, $f>0.9$ Hz, it appears that the more submerged the cylinders are, the lower is the resulting dissipative capacity.

\begin{figure}
  \centerline{\includegraphics[]{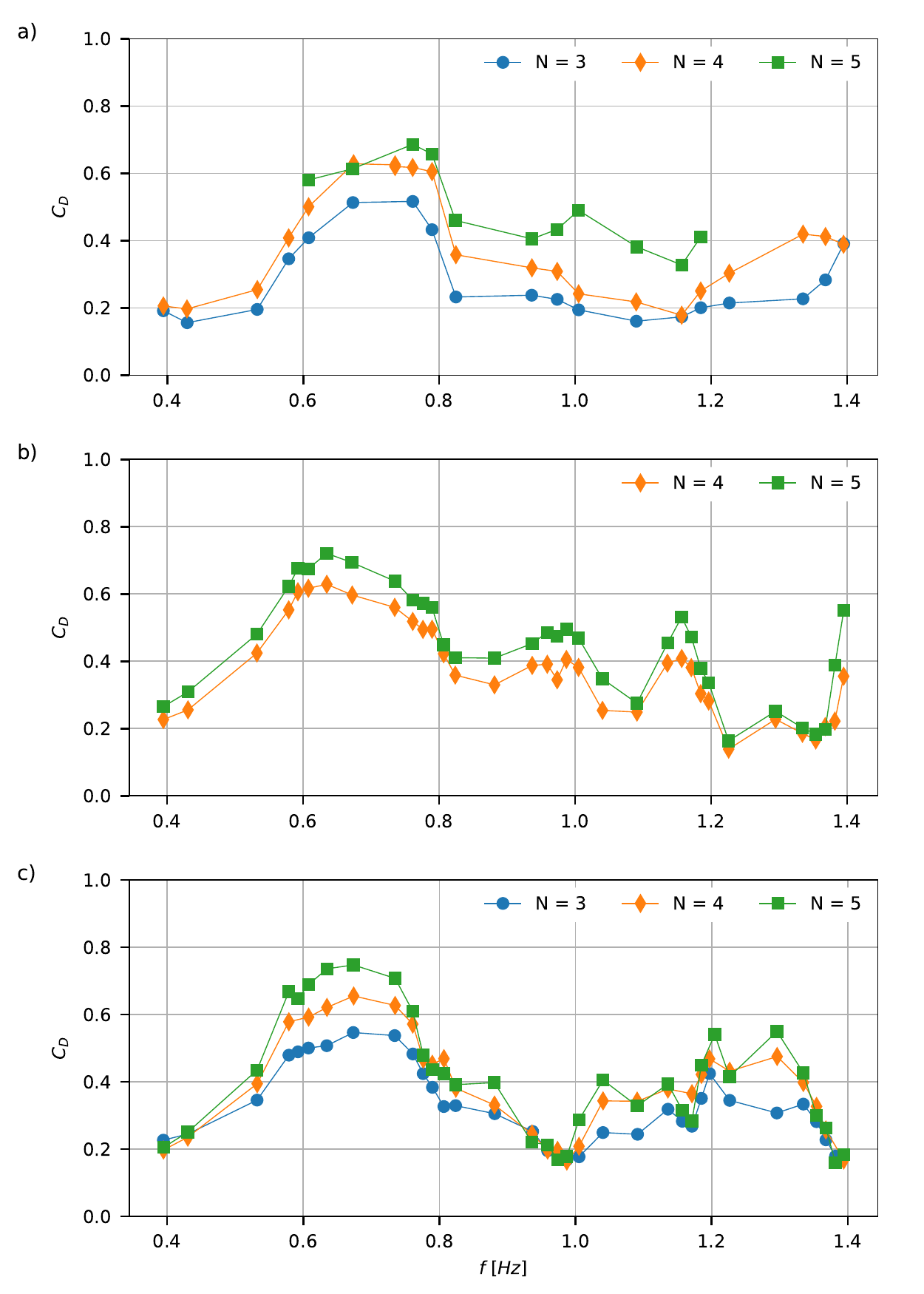}}
  \caption{Dissipation coefficient for various geometrical configurations and varying number of pendula: a) $L=2L^*$, b)$L=4L^*$ and c) $L=7L^*$.}
\label{fig:dissip_543}
\end{figure}

In Figure \ref{fig:dissip_543}, the role of the cylinder mutual distance $L$ is considered. Panels a), b) and c) depict the results for configurations with $L=2 L^*$, $4 L^*$ and $7 L^*$, respectively. Each line represents $C_D$ referred to different number of pendula ($N$). In general, for any $L$, energy losses increase with the number of cylinders. Moreover, the shape of the main dissipation bump in the $0.6-0.8 $ Hz band is very similar, independently of $L$. On the other hand, in the high frequency region, $f>0.9$ Hz, a different value of $L$ corresponds to a different response curve, and the shape of the response is unaffected by the number of cylinders. Specifically, taking $L=2 L^*$ as the baseline, an increase of the pendula spacing corresponds to the appearance of dissipation peaks at specific frequencies. Regardless of the number of pendula, for $L=4 L^*$, these peaks locate around $1.0$ Hz and $1.15$ Hz. The similarities between the 3 and 4 cylinder line are also clear for $L=7 L^*$.

\begin{figure}
  \centerline{\includegraphics[]{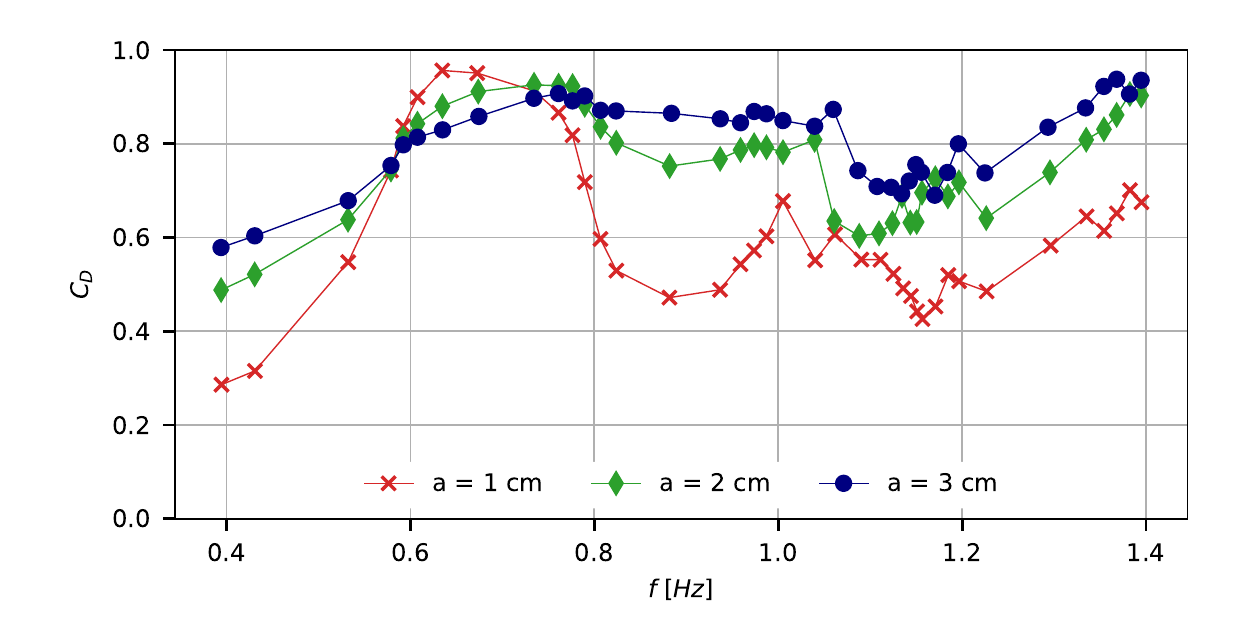}}
  \caption{Dissipation coefficient for 11 cylinders, L = 0.24m, h=0.45m and three different incoming wave amplitudes.}
\label{fig:diss_11cil_ampdiverse}
\end{figure}

Wave amplitudes also play a role: in figure \ref{fig:diss_11cil_ampdiverse}, the response of the 11-cylinder configuration is taken as reference, for $L=2L^*$, a $0.45$ m water depth, excited by small amplitude waves $a=1$ cm, as already depicted in figure \ref{fig:dissip_543}. The other two lines refer to wave amplitudes $a = 2$ cm and $a = 3$ cm, respectively. In general, larger amplitude waves correspond to a flatter response curve. Within the $0.6-0.8 $ Hz band, the dissipation decreases with the wave amplitude. On the contrary, higher energy dissipation takes place for shorter and longer waves. Note that, especially for $a=3$ cm, but also in some cases with $a=2$ cm, at least the first two or three cylinders emerged above water during the passage of the wave troughs.

\subsection{Off-shore reflection}\label{ss:res_ref}

While energy dissipation appears smooth across the tested frequency range, the amount of energy reflected back off-shore is concentrated at relatively sharp peaks.

\begin{figure}
  \centerline{\includegraphics[]{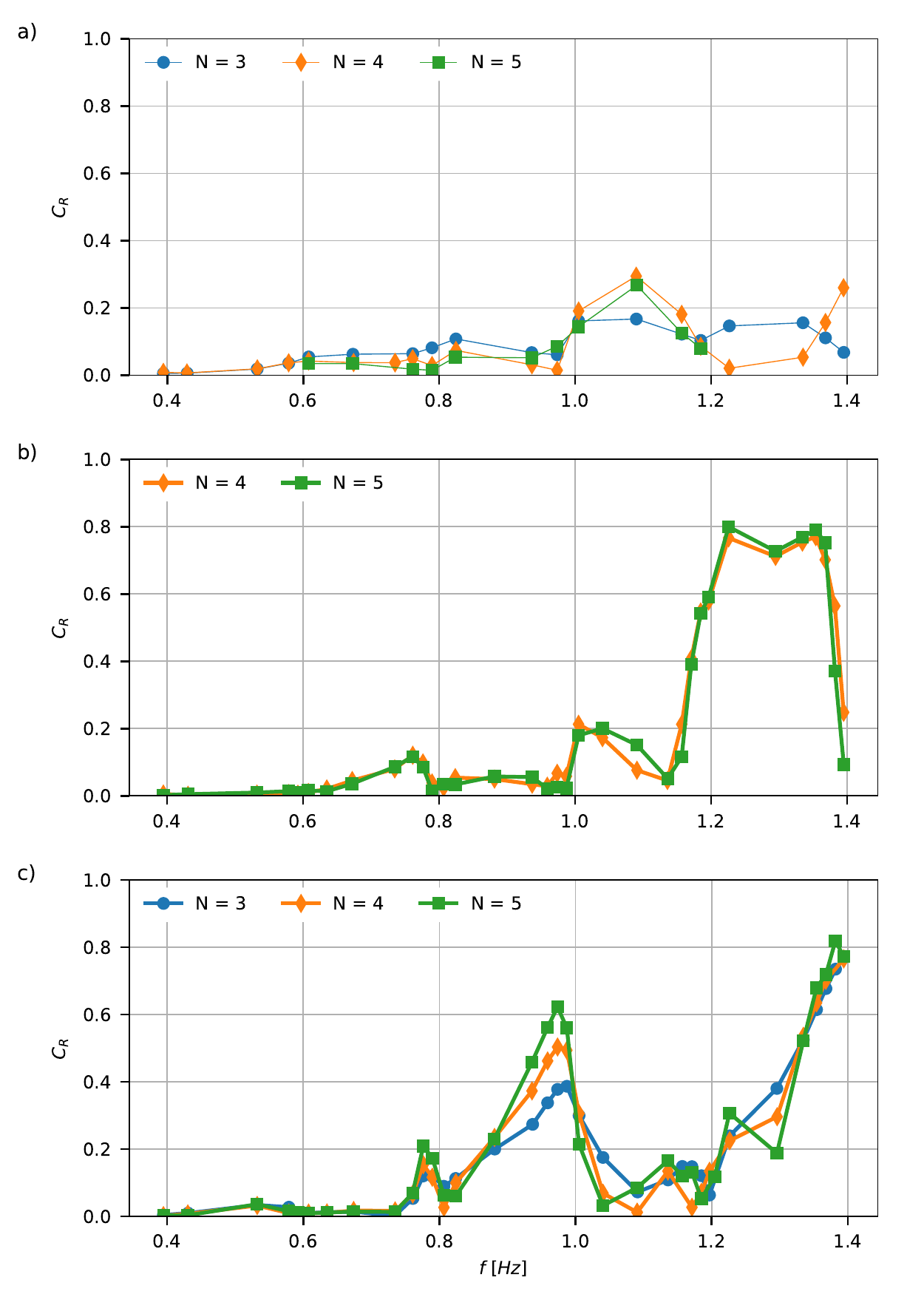}}
  \caption{Reflection coefficients for the same configurations as in  figure \ref{fig:dissip_543}.}
\label{fig:refl_543}
\end{figure}

Figure \ref{fig:refl_543} shows $C_R$ for different configurations, namely $L=2L^*$ in panel a), $L=4L^*$ in panel b) and $L=7L^*$ in panel c). As in figure \ref{fig:dissip_543}, each panel depicts the response of 3, 4 and 5 cylinders.
In general, the device is scarcely capable of reflecting back the incoming energy, except at specific frequencies, where sharp peaks appear. These peak positions do not vary with the number of cylinders. Only the height of the peaks, \emph{i.e.} the amount of reflected energy, appears slightly larger if the cylinders array is longer.
Each geometrical configuration displays its own peculiar frequency response. Increasing the cylinder spacing, the high reflection ($C_R>0.4$) peaks migrate to lower frequencies.

\begin{figure}
  \centerline{\includegraphics[]{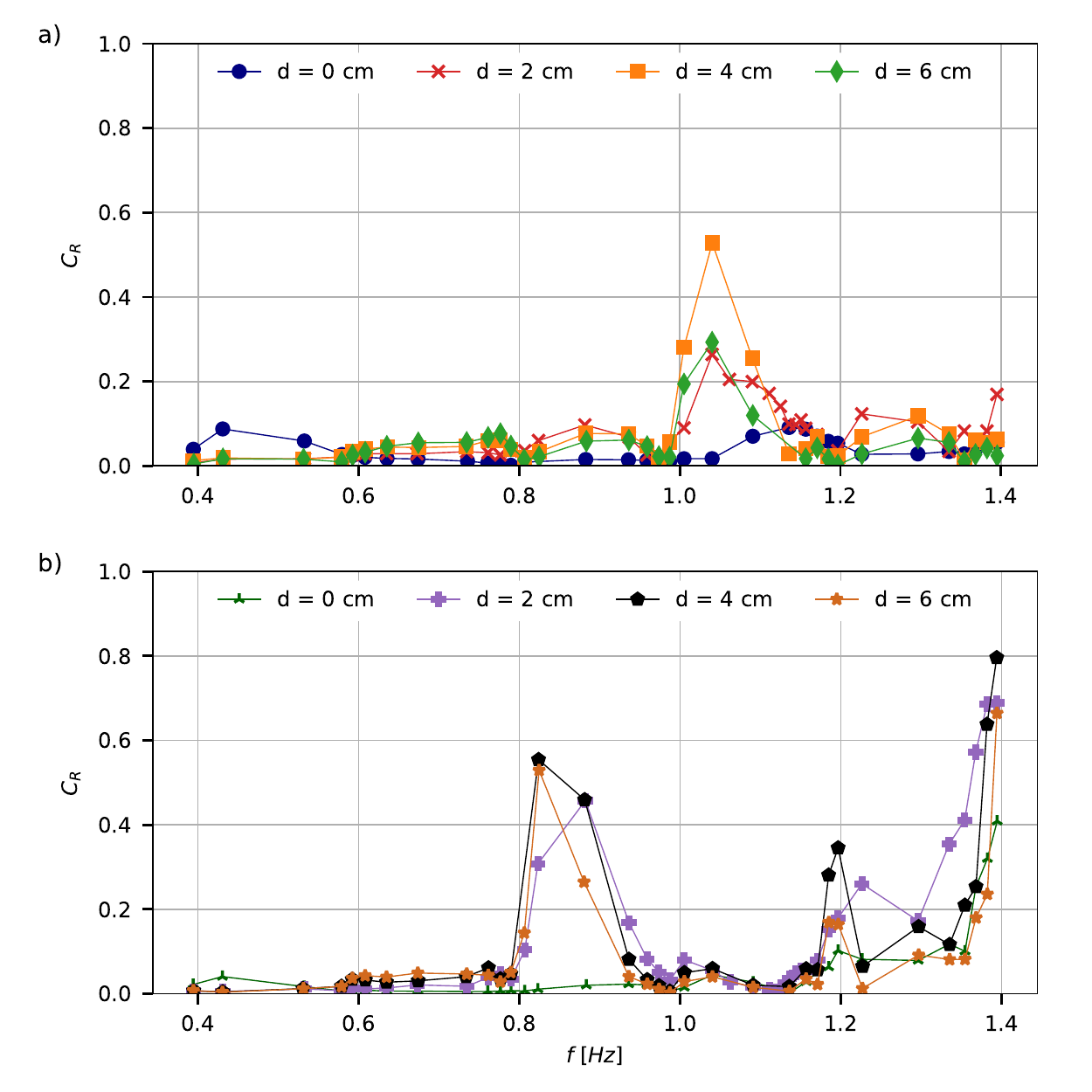}}
  \caption{Reflection coefficients for 11 cylinders, h=0.45m, a=1cm for two different spacings: a) L=0.24m, b) L=0.35m.}
\label{fig:refl_Lfix}
\end{figure}

To investigate the effects of pendula submergence, figure \ref{fig:refl_Lfix} documents a configuration (11 cylinders, $L=3L^*$) not considered in previous plots. This is useful for the interpretation of the reflection patterns, and is discussed in section \ref{ss:dis_L}. Reflection was almost negligible if the pendula rest position corresponded to the still water level, $d=0$. If the device was sufficiently submerged, the reflection coefficient presented localized peaks, similar to those observed in figure \ref{fig:refl_543}. Except in a small interval around $f=1.2$ Hz, where the behaviour is not sufficiently clear, both locations and intensities of the mid-frequency peak (around $0.85$ Hz) and the high frequency peak ($f\geq1.4$ Hz) are likely independent on the level of immersion.

\subsection{Nearshore transmission}
\label{ss:trasm}
On the nearshore side of the device, transmitted waves were measured using a single gauge. While dissipation and reflection measurements are important to understand the fluid dynamics around the pendula, transmission returns the residual energy of the incoming wave after the interaction with the device, providing an overall indication of the effectiveness of the system. Smaller transmission coefficients correspond to better energy wave attenuation.

\begin{figure}
  \centerline{\includegraphics[]{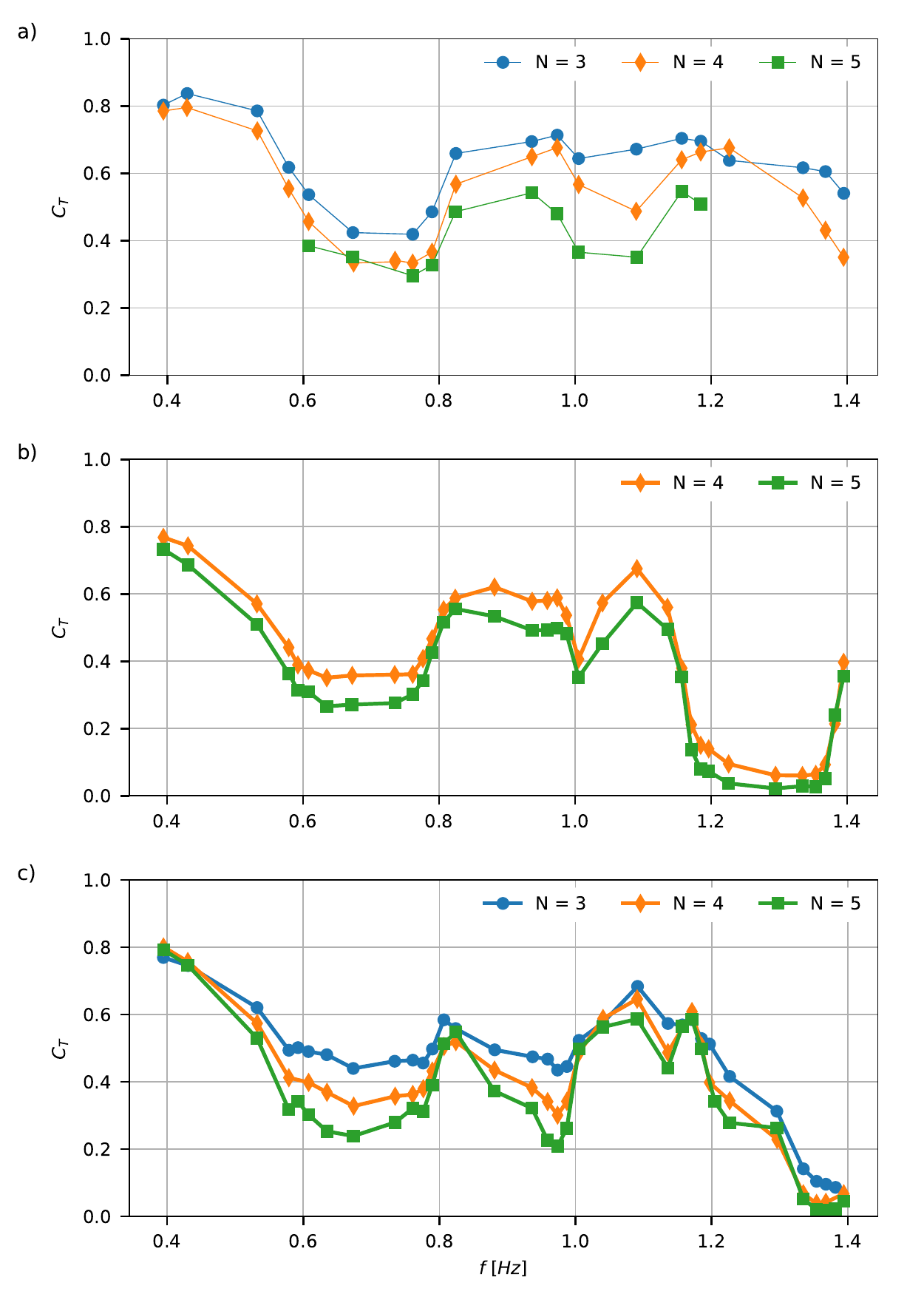}}
  \caption{Transmission coefficients for the same configurations of figure \ref{fig:dissip_543}.}
\label{fig:trasm_543}
\end{figure}

\begin{figure}
  \centerline{\includegraphics[]{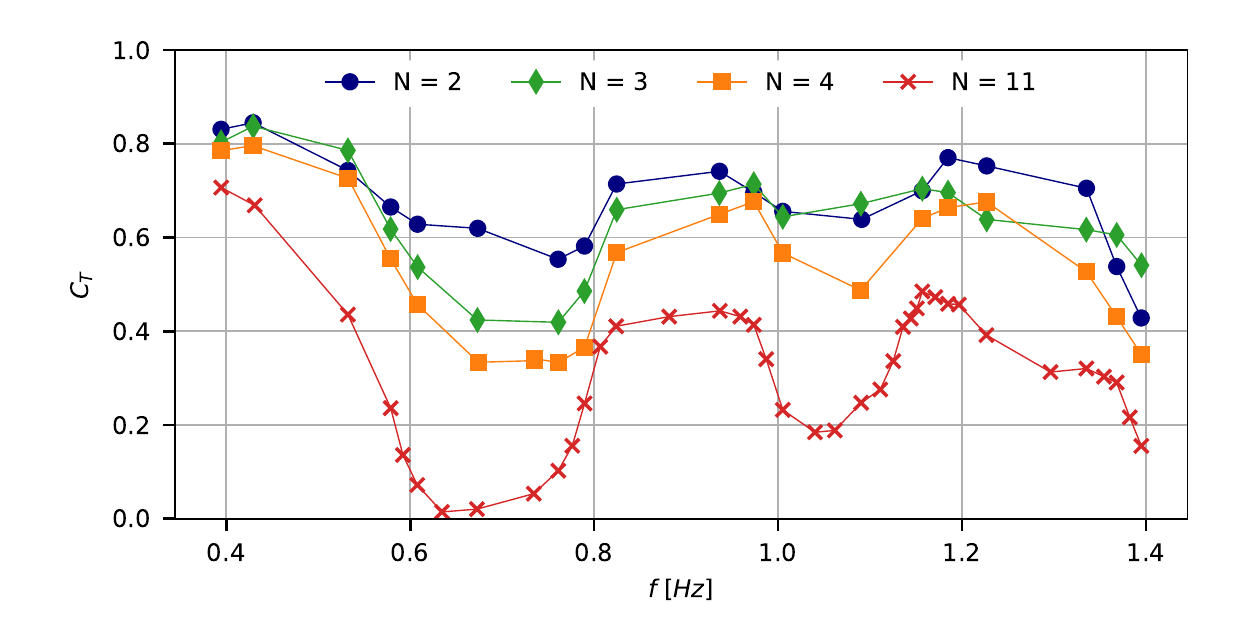}}
  \caption{Transmission coefficients for 11 cylinders, h=0.45m, a=1cm, L=2$L^*$.}
\label{fig:serie_vedarmo}
\end{figure}

In figure \ref{fig:trasm_543}, the transmission coefficients $C_T$ are plotted as function of the incident wave frequency for the same configurations depicted in figure \ref{fig:dissip_543} (dissipation) and figure \ref{fig:refl_543} (reflection).
Similarly, for the most investigated configuration $L=2L^*$, we present in figure \ref{fig:serie_vedarmo} the transmission results for varying number of cylinders, which can be compared to figure \ref{fig:dissip_11cil}, panel a), (dissipation) and figure \ref{fig:refl_Lfix} (reflection).

\section{Discussion}
\label{sec:discussion}

\subsection{Cylinders mutual distance}\label{ss:dis_L}
We have seen that the ability of the device to reflect the energy is not uniform, and concentrated in narrow bands at well-defined frequencies, which depend on the spacing $L$.
Let assume that the dynamics of our idealized device is ruled by the following condition, also known as exact resonance, see \cite{nazarenko2011wave}:  
\begin{subequations}
  \begin{equation}
    \label{eq:freqres}
    \sum_{i=1}^{N} s_i f_i = 0,
  \end{equation}
  \begin{equation}
    \label{eq:wvnres}
    \sum_{i=1}^{N} s_i {\rm sign}(k_i) k_i = 0,
  \end{equation}
\end{subequations}
with $s_1 = \pm1$, $k_i>0$ and ${\rm sign}(k_i)=\pm 1$ indicates the direction of the wave (+1 for waves travelling towards the shore and -1 for waves travelling in the opposite direction).  
If we neglect the pendula motion, their frequency is zero ($f_c=0$) and the associated wavenumber is determined by the spacing $L$, \emph{i.e.} $ k_L=2\pi/L$. The water waves instead obey the linear dispersion relation \citep{Whi74}
\begin{equation}\label{eq:disprel}
  4\pi^2 f^2 = \gi k \tanh{kh}.
\end{equation}

Now, with $N=3$, \eqref{eq:freqres} is satisfied only if $f_1=f_2=f$ with $s_1=1$, $s_2=-1$ and $f_3=f_c=0$. From the wavenumber condition \eqref{eq:wvnres}, if we take two water waves travelling in opposite directions, \emph{i.e.} $k_1=k$ and $k_2=-k$, we have the simplest Bragg resonance:
\begin{equation}\label{eq:bragg1_scatt}
  2 k = k_L,
\end{equation}
all other combinations being meaningless. In the same way, for $N\geq3$, admitting only the existence of two waves having the same frequency and travelling in opposite directions, we have the generalized Bragg resonance condition
\begin{equation}\label{eq:braggn_scatt}
  2 k = n k_L
\end{equation}
with $n \in \mathbb{N}$. For any $n$, this condition states that there is a possible energy transfer from the incoming wave, $k$, to the backward travelling one, $-k$, \emph{i.e.} the device could act as a reflector.

In a wave flume experiment, even dealing with small amplitude regular waves, we cannot neglect the existence of higher harmonics. The wavemaker sinusoidal motion used in this study causes the release of spurious components, the bigger ones being second harmonics that have the same order of magnitude of the Stokes bound modes, but travel at their own speed, \emph{i.e.} their wavenumber is determined by \ref{eq:disprel} \citep[see e.g.][]{schaffer96}. Also, \citet{Grue1992} and \citet{Chaplin1984}, in the case of a single submerged cylinder, observed the generation of the second and third harmonics. Therefore, we can expect that similar effects take place also for waves interacting with the pendula array, and admit that, if $f$ is the frequency of the fundamental wave, higher harmonics may be travelling in the flume, satisfying the dispersion relation
\begin{equation}\label{eq:disprelm}
  4\pi^2 m^2 f^2 = \gi k^{\left(m\right)} \tanh{h k^{\left(m\right)}},\;m\in\mathbb{N}
\end{equation}
 and playing a role in the dynamics. Including these waves into the resonance conditions, we find interactions of the type:
\begin{equation}
  \label{eq:2nd_scattn}
  2 k + k^{\left(2\right)} = n k_L,
\end{equation}
and a number of other conditions. These include for example the Bragg resonances of the second harmonic, and other situations, which require that the wavemaker-generated second harmonic be large enough. This is not our case, at least in small amplitude tests. On the other hand, the condition \eqref{eq:2nd_scattn} takes into account second harmonics travelling back to the wavemaker, after being ``generated'' from the interaction of the carrier with the pendula lattice.

\begin{figure}
  \centerline{\includegraphics[]{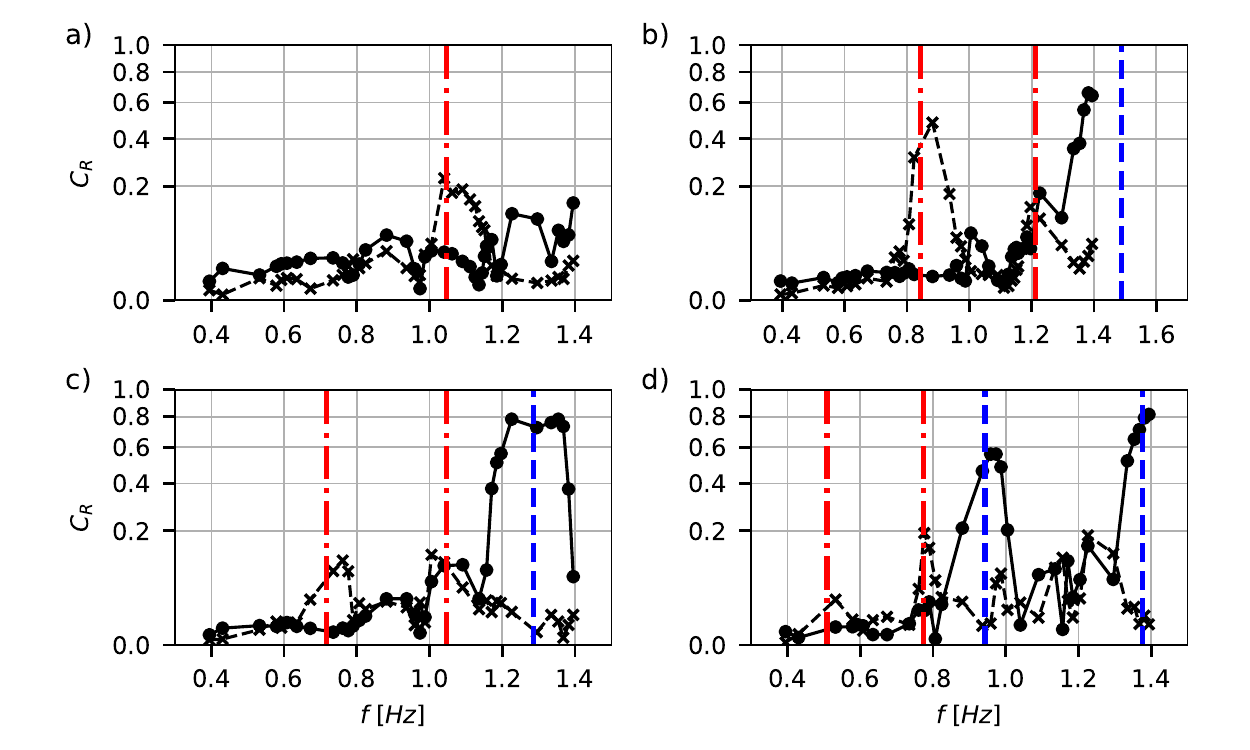}}
  \caption{Reflection coefficients for the fundamental (dots, solid) and second harmonic (crosses, broken line). a) L=2$L^*$, $N = 11$ , b) L=3$L^*$, $N=11$, c) $N=5$,  L=4$L^*$, d) $N=5$, L=7$L^*$. Blue: Bragg scattering frequencies \eqref{eq:braggn_scatt}, red: second harmonic generation \eqref{eq:2nd_scattn}.
  Note that ordinate axis scales with a quadratic power law.
  }
\label{fig:refharms}
\end{figure}

Table \ref{tab:braggs} summarizes the first solutions of the generalized Bragg condition \eqref{eq:braggn_scatt} and second harmonic reflection \eqref{eq:2nd_scattn}. Around these frequencies, the most significant peaks in the reflection coefficients, presented in section \ref{ss:res_ref}, can be observed. In order to facilitate the interpretation, we reproduce some of the above results in Figure \ref{fig:refharms}. Here, the reflection coefficients are given for the fundamental frequency $f$ and the second harmonic ($2f$), both with reference to the incoming energy of the fundamental. The four panels reproduce the four geometrical configurations, with different $L$ values. The number of cylinders is also different, but it is not significant for the present analysis. Around the blue lines (Bragg scattering conditions), the reflection is relevant and almost all the energy is transferred to the reflected fundamental, testifying a Bragg-like scattering effect. Around the red lines, the reflected second harmonic is strongly enhanced, while the fundamental is very small. This confirms the predictions of equation \eqref{eq:2nd_scattn}.
As pointed out in the review by \citet{Patil2021}, this second harmonic generation effect is a typical non-linear phenomena linked to the geometrical configuration of the device \citep[see also][]{Guo2019, Raju2022}. Indeed, in our tests this behavior can be observed even for very small amplitude incident waves (\textit{i.e.} linear regime).
Notice that the magnitude of the reflection peaks appears to decay with larger $L$, as the resonant frequencies become smaller. However, in the longer wave region, close to the cylinder natural frequency, viscous dissipation mechanisms are dominant, masking the actual magnitude of scattering response.

\begin{table}
  \begin{center}
  \begin{tabular}{cccccc}
    $L\;\;\;$   &
    Eq. \eqref{eq:braggn_scatt} $n=1\;\;\;$ &
    Eq. \eqref{eq:braggn_scatt} $n=2\;\;\;$ &
    Eq. \eqref{eq:2nd_scattn} $n=1\;\;\;$ &
    Eq. \eqref{eq:2nd_scattn} $n=2\;\;\;$ \\
     & [Hz]  & [Hz] & [Hz] & [Hz] \\
    2$L^*\;\;\;$ &            -$\;\;\;$ &            -$\;\;\;$ &                       1.046$\;\;\;$ &                            -$\;\;\;$\\
    3$L^*\;\;\;$ &        1.488$\;\;\;$ &            -$\;\;\;$ &                       0.843$\;\;\;$ &                        1.213$\;\;\;$ \\
    4$L^*\;\;\;$ &        1.286$\;\;\;$ &            -$\;\;\;$ &                       0.717$\;\;\;$ &                        1.046$\;\;\;$ \\
    7$L^*\;\;\;$ &        0.943$\;\;\;$ &        1.376$\;\;\;$ &                       0.509$\;\;\;$ &                        0.774$\;\;\;$
  \end{tabular}
  \caption{Scattering frequencies for (fixed) cylinder array with mutual distance L (rows). First two exact solutions of \eqref{eq:braggn_scatt} (Bragg scattering) and \eqref{eq:2nd_scattn} (second harmonic generation).}
  \label{tab:braggs}
  \end{center}
\end{table}

\subsection{Number of cylinders}

Very large dissipation occurs, for any configuration, in the same $0.6-0.8$ Hz band.
For the submerged device ($d>0$ in Figure \ref{fig:dissip_11cil} b) ), the behavior of the resonance frequency band doesn't change with the water level, and, as seen in subsection \ref{sub:freq_pend}, its position depends mainly on the characteristics of the single pendula.

According to equation \eqref{eq:equilibrium}, we assume that the system is closed. Therefore, the dissipation results must be interpreted as total energy loss, which not only depends on the energy exchanges in water but also between water and air, and the damping of the flume bottom and walls. The motion of the pendula is also damped by viscosity.

Overall, it appears that dissipation depends on the single cylinder behavior. Indeed, energy loss increases with the number of elements (see Figures \ref{fig:dissip_11cil} a) and \ref{fig:dissip_543}), while, fixing all other variables, the response curves are similar.
In summary, increasing the number of cylinders leads to a decrease of the transmitted energy to the shore in the whole frequency range (see section \ref{ss:trasm}). If the number of cylinders is sufficiently large, all the incoming energy, in a frequency interval around the resonant frequency, can be dissipated. The final effect is thus similar to the formation of a bandgap, as can be observed in figure \ref{fig:serie_vedarmo} in the region $0.6-0.8 $ Hz.

\subsection{Submergence of the cylinders and dependence on wave amplitude}

In principle, it is not possible to fully decouple the effects of pendula submergence and incoming wave amplitude, simply because for wave amplitudes larger than the baseline ($1$ cm), at least the first one or two cylinders were always exposed to air during the tests.

However, there are two clear effects that depend on the submergence of the pendula. The first relates to its dissipative capacity, as can be observed in Figure \ref{fig:dissip_11cil} b). For all fully submerged configurations, most of the dissipation occurs at frequencies which are just above the natural frequency of the pendula, but for $d=0$ (the cylinders are fully immersed only with the water at rest), the dissipation peak moves toward longer waves. This is consistent with previous studies \citep{Sergiienko2017} showing that, for long waves, an emerging device is more effective. Also, shorter waves lose significant energy only if the wave amplitude ($1 $ cm in this case) is comparable with the gap between the mean water level and the top of the cylinders. Thus, in the high frequency range (above $1$ Hz), less immersed devices are more effective. This is probably due to surface breaking effects which can become important for small submergence.

The second effect is visible in both panels of Figure \ref{fig:refl_Lfix}. As discussed, the peaks around $1.1$ Hz and $0.95$ Hz (panels a) and b), respectively) are due to second harmonic generation. If the cylinders are not fully immersed ($d=0$ lines), no refection peak is observed, meaning that the second harmonic generation mechanism is triggered only if the pendula are fully immersed.

Increasing the wave amplitude, while keeping the same mean water level, the frequency response does not change drastically. The curves flatten, with short waves losing more energy, but the high dissipation in the $0.6-0.8 $ Hz remains. Increasing wave amplitude leads to the onset of other mechanisms, such as wave breaking and other non-linear phenomena, thus making the behaviour of the device more complex than that expected from approximately linear interactions.

The above analysis captures almost all the features in the transmission and reflection spectra, but it does not explain the observed behaviour in the interval $1.2-1.3$ Hz. In this range, there is small but finite reflection in all configurations. We note that the frequency of $1.2$ Hz corresponds to twice the natural frequency of the pendulum, but is also close to the channel first transverse mode \citep{Radhakrishnan2007}. During the tests with frequencies close to $1.2$ Hz we actually observed the excitation of non negligible transverse modes, causing the first one or two pendula to rotate on the horizontal plane.

\section{Conclusions}
\label{sec:concl}

This work draws its motivation from the attempt to address the problem of coastal erosion. The idea is to exploit the concept of metamaterial wave control, building a device to mitigate surface gravity waves. In particular, the objective is to exploit both local resonance effects and Bragg scattering from a periodic array of structures to obtain attenuation in a wide frequency range. 
Following numerical results presented by \citet{DeVita2021meta}, we experimentally investigated a first prototype in a two-dimensional wave flume using a 1D array of submerged and reversed cylindrical pendula. We performed single-frequency wave tests over a wide range of frequencies, aiming to maintain the system in the linear regime as far as possible. Experiments were carried out varying various parameters, including the mutual distance $L$, the number of pendula, their submergence and the amplitude of the incoming waves.

Results demonstrated the feasibility of the concept, and that wave attenuation can be significant even using a limited number of cylinders and without any particular optimization, see \citet{krushynska2017coupling}.  
In experiments, wave attenuation reached 80\% over a large range of frequencies in the case of 11 cylinders. In this configuration, the submerged device ceased to be effective only in attenuating very long waves.

Analysis of results allowed to assess how the two leading mechanisms (reflection and dissipation) contribute to wave attenuation. The main discriminant was the submergence of the device. If the pendula were fully immersed, dissipation induced a broad band-gap around a resonance frequency that depends on the characteristics of the pendula, while non-fully immersed pendula tended to dissipate both longer and shorter waves. It remains to be understood in this case if the natural frequency of the pendula played a role.
The scattering behaviour was also different if the pendula were immersed or not. Only in the former case, some of the incoming energy was prevented from propagating through the array and was reflected back to sea. This occurred in narrow bands close to four and five wave interaction frequency conditions (including harmonics), which depend on the geometry of the whole structure, with Bragg scattering being only the simplest mechanism.

Regarding the feasibility of appropriately designing the cylinder array, an optimal device should prevent or mitigate the propagation of waves to the shore for a particular wave climate.
As in the case of elastic metamaterials, appropriate choice of geometrical parameters can allow to tune band-gap frequencies to desired values, whether they be dissipation or reflection related.
The dissipation band can be adjusted by appropriate choice of pendulum parameters, since it is mainly related to their natural oscillation frequency. It is expected that, at prototype scale, dissipation will vary compared to that observed in the lab. Thus, in order to move to a proof of concept device, it is necessary to properly understand the scaling of viscous effects and verify if they remain relevant for real sea conditions.

We have also seen that when pendula remain completely submerged, wave scattering is relevant around frequencies that can be predicted by Bragg scattering relations. Further investigations are required to understand how the amplitude or the width of the observed frequency bands vary, but we can expect these phenomena to remain consistent as the system is rescaled. This implies that it should be possible to tune the position of the reflection bands in the wave spectrum by modifying system parameters, \emph{e.g.} the mutual distance of the pendula.
Overall, the presented design concept appears promising in the quest to realize a new class of multifrequency surface gravity wave absorbers with increased efficiency and limited environmental impact, to be applied \emph{e.g.} for the protection against beach erosion in practical scenarios.


\section*{Acknowledgements}
The authors wish to thank Costantino Manes and Carlo Camporeale for their valuable suggestions during the early stages of the work and for providing access to the laboratory, Gaetano F. Caminiti for his help in setting up the experimental apparatus, and Vito Jr. Battista for producing photographs and video material.
\section*{Funding}
This work has been funded by Compagnia di San Paolo Progetto d’Ateneo - Fondazione San Paolo \textit{Metapp}, n. CSTO160004. and Proof of concept \textit{Metareef} grant. MO, FB, FDL and ML acknowledge the EU, H2020 FET Open \textit{Bio-Inspired Hierarchical MetaMaterials} (grant number 863179).
\section*{Declaration of interests}
The authors report no conflict of interest.

\section*{Author ORCID}
\begin{itemize}
  \item[] M. Lorenzo, https://orcid.org/0000-0001-8330-2327
  \item[] P. Pezzutto, https://orcid.org/0000-0002-5415-8191
  \item[] F. De Lillo, https://orcid.org/0000-0002-1327-695X
  \item[] F. M. Ventrella, https://orcid.org/0000-0001-7367-1444
  \item[] F. De Vita, https://orcid.org/0000-0001-8616-269X
  \item[] F. Bosia, https://orcid.org/0000-0002-2886-4519
  \item[] M. Onorato, https://orcid.org/0000-0001-9141-2147
\end{itemize}

\bibliographystyle{jfm}
\bibliography{metajfm22}

\end{document}